\documentclass[conference]{IEEEtran}
\IEEEoverridecommandlockouts
\usepackage{cite}
\usepackage{amsmath,amssymb,amsfonts}
\usepackage{algorithmic}
\usepackage{graphicx}
\usepackage{romannum}
\usepackage{textcomp}
\usepackage{mathtools}
\usepackage{xcolor}
\def\BibTeX{{\rm B\kern-.05em{\sc i\kern-.025em b}\kern-.08em
    T\kern-.1667em\lower.7ex\hbox{E}\kern-.125emX}}
\begin{document}

\title{Training DNA Perceptrons via Fractional Coding}
\author{\IEEEauthorblockN {Xingyi Liu and Keshab K. Parhi} \\
\vspace{-1em}
\IEEEauthorblockA{Department of Electrical and Computer Engineering \\
University of Minnesota, Minneapolis, MN 55455, USA \\
Email: liux3138@umn.edu, parhi@umn.edu}
}
\maketitle

\begin{abstract}
This paper describes a novel approach to synthesize molecular reactions to train a perceptron, i.e., a single-layered neural network, with sigmoidal activation function. The approach is based on {\em fractional coding} where a variable is represented by two molecules. The synergy between fractional coding in molecular computing and stochastic logic implementations in electronic computing is key to translating known stochastic logic circuits to molecular computing. In prior work, a DNA perceptron with bipolar inputs and unipolar output was proposed for inference. The focus of this paper is on synthesis of molecular reactions for {\em training} of the DNA perceptron. A new {\em molecular scaler} that performs multiplication by a factor greater than $1$ is proposed based on fractional coding. The training of the perceptron proposed in this paper is based on a {\em modified} backpropagation equation as the exact equation cannot be easily mapped to molecular reactions using fractional coding.
\end{abstract}

\begin{IEEEkeywords}
Molecular Computing, Fractional Coding, Perceptron, Backpropagation, Molecular Scaler, DNA Computing.
\end{IEEEkeywords}

\section{Introduction}

Since the pioneering work on DNA computing by Adleman, there has been growing interest in DNA computing for signal processing and machine learning functions \cite{adleman1994molecular}. For example, several logic functions and simple arithmetic operations have been simplemented {\em in vitro} using bimolecular reactions~\cite{zhang2009control,soloveichik2010dna,qian2011neural}. In recent work, {\em fractional coding} has been introduced in \cite{salehi2015markov,salehi2016chemical} for molecular implementations of Markov chains and for computing polynomials using Bernstein expansion. The synergy between fractional coding in molecular computing and stochastic logic in electronic computing was first established in~\cite{salehi2018computing}. This synergy is significant as it enables every stochastic logic circuit to be translated to a molecular circuit. Based on prior stochastic logic circuit implementations in \cite{parhi2016computing}, several complex mathematical functions and perceptrons for inference applications were implemented using molecular computing based on fractional coding \cite{salehi2018computing}. A molecular radial basis function kernel for support vector machines was also presented in \cite{liu2019computing} based on fractional coding.

Inspired by stochastic logic for electronic computing~\cite{gaines1967stochastic}, a value $X$ can be represented by two molecules: a $1$-molecule ($X_1$) and a $0$-molecule ($X_0$). All $1$-bits are grouped to form the $1$-molecule and all $0$ bits to the $0$-molecule ~\cite{salehi2018computing}. Unlike electronic stochastic logic, the molecular circuits do not suffer from correlation effects and the computation is deterministic. Each value can be either unipolar or bipolar. In a unipolar representation, the variable is bounded between $0$ and $1$. In a bipolar representation, the dynamic range of the variable is between $-1$ and $1$.

In the unipolar format, the value of a variable is given by:

\begin{equation}
\begin{split}
x=\frac{[X_1]}{[X_0]+[X_1]}\nonumber
\end{split}
\end{equation}
where $[X_0]$ and $[X_1]$ correspond to the concentrations of the assigned molecular types $X_0$ and $X_1$, respectively. In the bipolar format, the value of a variable is given by:

\begin{equation}
\begin{split}
x=\frac{[X_1]-[X_0]}{[X_0]+[X_1]}\nonumber
\end{split}
\end{equation}
where $[X_0]$ and $[X_1]$ are defined as before.

\begin{figure}[h]
\vspace{0em}
\centering
\resizebox{0.47\textwidth}{!}{%
\includegraphics{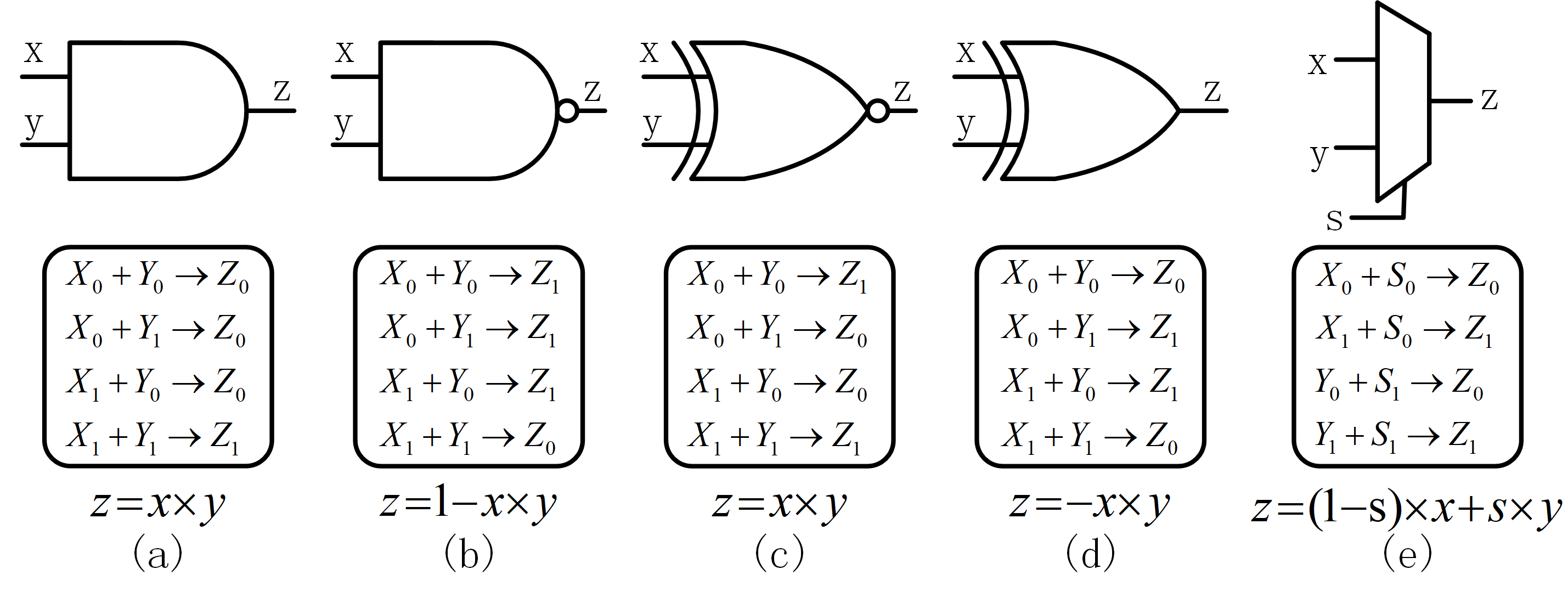}}
\caption{Basic molecular units as described in \cite{salehi2018computing}. (a) The {\tt Mult} unit with unipoar inputs and output. (b) The {\tt NMult} unit with unipolar inputs and output. (c) The {\tt Mult} unit with bipolar inputs and output. (d) The {\tt NMult} unit with bipolar inputs and output. (e) The {\tt MUX} unit with unipolar/bipolar $x$, $y$ and $z$ and unipolar $s$.}
\label{base}
\vspace{-1em}
\end{figure}

Molecular reactions for the {\tt Mult} and {\tt NMult} units that perform multiplication using unipolar or bipolar representation as well as {\tt MUX} unit that computes scaled addition have been presented in \cite{salehi2018computing} and are illustrated in Fig.~\ref{base}.

The four molecular reactions shown in Fig.~\ref{base}(a) compute $z=x\times y$, all in unipolar format. We refer to this molecular unit as {\tt Mult} unit. Fig.~\ref{base}(b) shows the molecular reactions that compute $z=1-x\times y$ in unipolar format, referred to as {\tt NMult} unit. Figs.~\ref{base}(c) and (d) illustrate the implementations of \texttt{Mult} unit and \texttt{NMult} unit with bipolar inputs and output, respectively. The bipolar \texttt{Mult} unit performs multiplication in the bipolar format. The bipolar \texttt{NMult} unit computes $z=-x\times y$, all in bipolar format. The {\tt MUX} unit that performs scaled addition is shown in Fig.~\ref{base}(e). This unit computes $z=(1-s)\times x+s\times y$. Notice that $x$, $y$ and $z$ can be in the unipolar format or bipolar format, but $s$ must be in unipolar format.

Stochastic logic implementations of complex functions such as exponential, logarithmic and trigonometric functions were presented in \cite{parhi2016computing}. By translating the stochastic logic gates to molecular reactions, the authors in~\cite{salehi2018computing} presented molecular implementations of these functions and perceptron. This paper presents an approach for the synthesis of molecular reactions for {\em training} a perceptron which are then mapped to DNA strand displacement (DSD) reactions~\cite{soloveichik2010dna}. A new {\em molecular scaler} that multiplies a protein by a value greater than $1$ using fractional coding is also presented in this paper. The performance of training a DNA perceptron with $3$ bipolar inputs and $1$ bipolar bias is presented in this paper.

This paper is organized as follows. Section \Romannum{2} presents a review of the perceptron with sigmoidal activation function. Molecular reactions of forward computation and backpropagation for a perceptron are, respectively, presented in sections \Romannum{3} and \Romannum{4}. Section \Romannum{5} presents the experimental results of a simple perceptron using DNA.

\section{Perceptron with Sigmoidal Activation Function}
\begin{figure}[htbp]
\vspace{-1em}
\centering
\resizebox{0.40\textwidth}{!}{%
\includegraphics{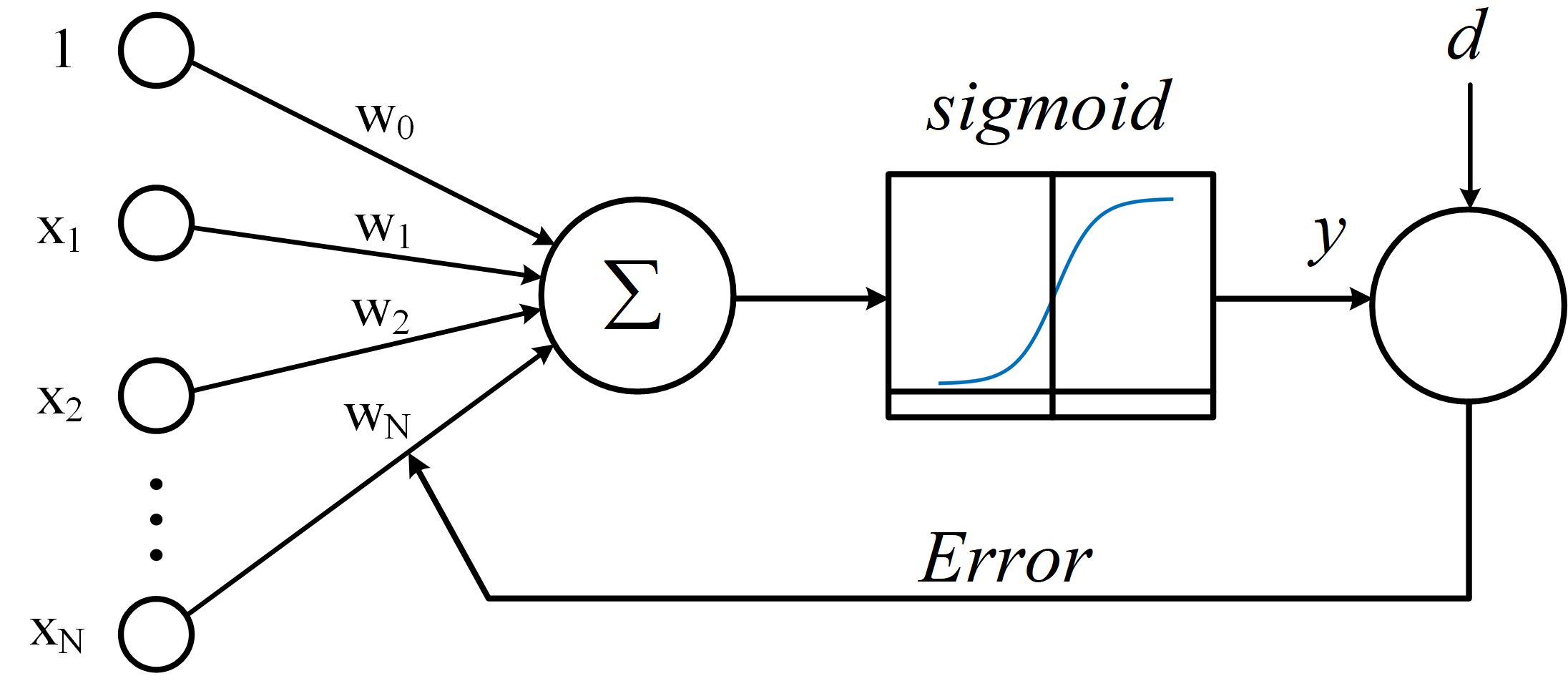}}
\caption{A typical perceptron that can apply sigmoid function to the weighted sum and update weights according to the distance between actual output $y$ and desired output $d$.}
\label{per}
\vspace{-1em}
\end{figure}

In machine learning, a perceptron is typically used for supervised learning of binary classifiers. A binary classifier is a function which can decide whether or not an input, represented by a vector of numbers, belongs to some specific class~\cite{freund1999large}. A perceptron with sigmoidal activation function shown in Fig.~\ref{per} can also be used in regression applications and is used as a basic unit for multi-layer neural networks.

The output of the perceptron is given by:

\begin{equation}
\begin{split}
y=sigmoid(\sum_{i=0}^N w_ix_i) \nonumber
\end{split}
\end{equation}
where $x_i$ and $w_i$ represent input and corresponding weight, respectively. The paameter $x_0$ is set to $1$ and $w_0$ represents the bias. In machine learning, backpropagation is widely used in the training of feedforward neural networks for supervised learning~\cite{rumelhart1986learning}. Backpropagation can efficiently compute the gradient of the loss function with respect to the weights of the network. In this paper, we define the loss function as $E=\frac{1}{2}(y-d)^2$ where $y$ is the actual output of the perceptron and $d$ represents the desired output for a given input vector $\textbf{x}$. Then the derivative of the loss function in terms of the weights $\frac{\partial E}{\partial w_i}$ is computed where $w_i$ is the $i^{th}$ weight for $i=0, 1, 2, \cdots N$. Each weight should be adapted to minimize the loss function as:

\begin{equation}
\begin{split}
w_{i,new}=w_i-\alpha \frac{\partial E}{\partial w_i} \nonumber
\end{split}
\end{equation}
where $\alpha$ represents the learning rate and is assumed to be a constant.

\section{Molecular Reactions of Forward Computation in Perceptron}
This section presents the molecular implementation of the forward computation of a perceptron where the inputs and output are in bipolar format. This computation can be divided into two parts: computing the scaled inner product of the input vector and their corresponding weights ($\frac{1}{N}\sum_{i=1}^N x_iw_i$), and computing the sigmoid function. The molecular inner product was first introduced in~\cite{salehi2018computing} and later revisited in~\cite{liu2019molecular}. The molecular implementation of the inner product scaled by sum of the absolute weights was also proposed in~\cite{liu2019molecular}. But this method is not suitable for training a perceptron as the weights are trained continuously during the training process. This method has a format conversion in the last step that converts the bipolar output to unipolar output in order to improve the precision. But in this paper, this format conversion is removed since bipolar outputs are needed for the training process.

\subsection{Inner Products Scaled by the Number of Inputs}
\begin{figure}[htbp]
\vspace{-0.8em}
\centering
\resizebox{0.3\textwidth}{!}{%
\includegraphics{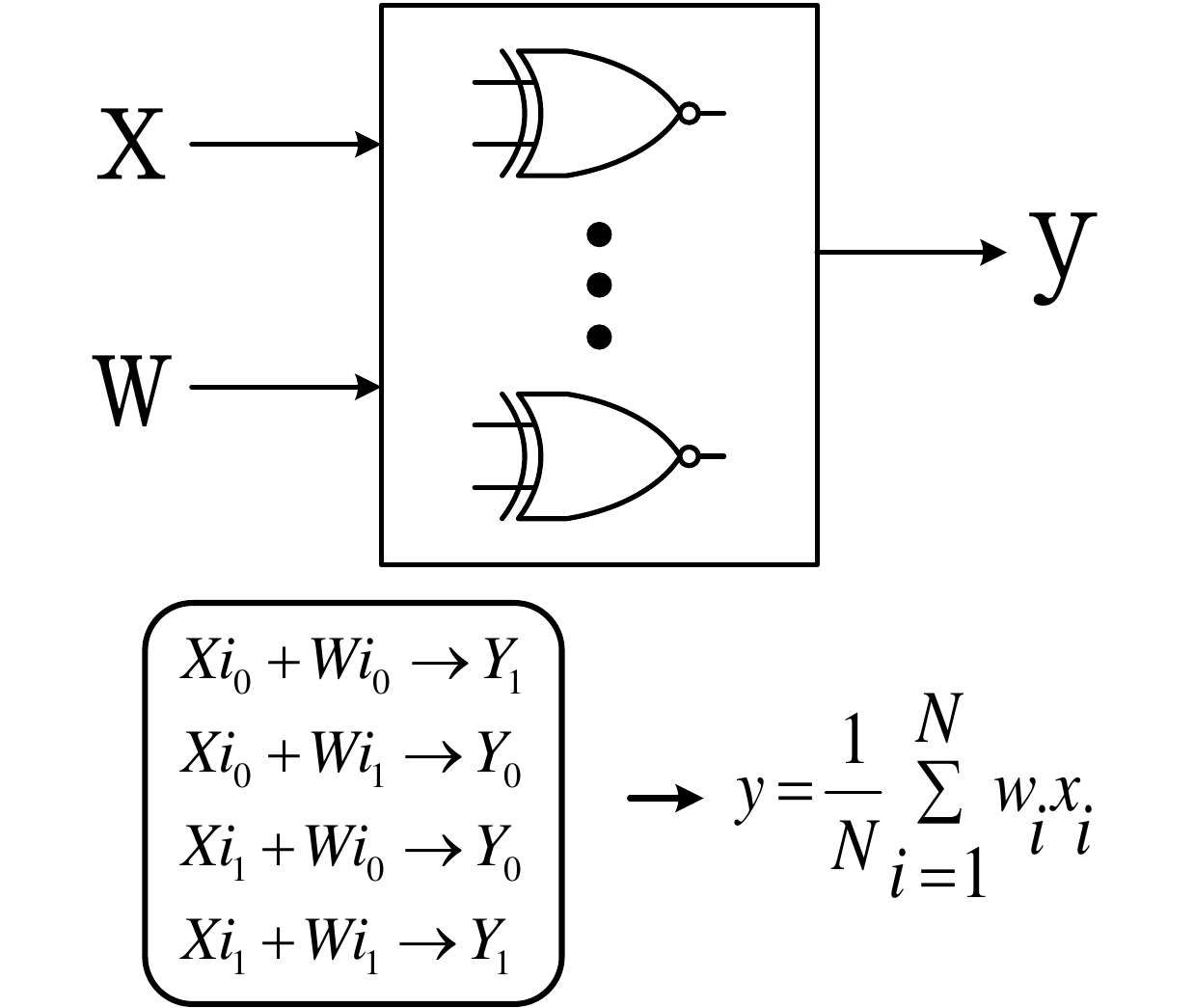}}
\caption{The inner product unit in~\cite{salehi2018computing}. This unit calculates $y=\frac{1}{N}\sum_{i=1}^{N}w_ix_i$, the scaled inner product of two input vectors $x$ and $w$ where each element is in bipolar fractional representation.}
\label{inner1}
\vspace{-1em}
\end{figure}

Given two bipolar input vectors, $x=\begin{bmatrix}x_1\\x_2\\:\\x_N\end{bmatrix}$ and $w=\begin{bmatrix}w_1\\w_2\\:\\w_N\end{bmatrix}$, the molecular implementation of inner product functions scaled by the number of inputs $N$ with $4N$ molecular reactions was proposed in~\cite{salehi2018computing}. As shown in Fig.~\ref{inner1}, the four reactions correspond to the \texttt{Mult} shown in Fig.~\ref{base}(c) with two corresponding inputs, $x_i$ and $w_i$. Fig.~\ref{inner1} also shows the proposed molecular reactions, where $i=1, 2, \cdots N$. Notice that $-1\leq x_i\leq 1$, $-1\leq w_i\leq 1$ must be guaranteed to meet the requirement of bipolar representation. Given $x_i=\frac{[Xi_1]-[Xi_0]}{[Xi_1]+[Xi_0]}$ and $w_i=\frac{[Wi_1]-[Wi_0]}{[Wi_1]+[Wi_0]}$ then $y=\frac{[Y_1]-[Y_0]}{[Y_1]+[Y_0]}=\frac{1}{N}\sum_{i=1}^{N}w_ix_i$. A proof of the functionality of the molecular inner product in Fig.~\ref{inner1} is described in Section S.5 of the Supplementary Information in \cite{salehi2018computing}.

\subsection{Implementation of Sigmoid Functions in Bipolar Format}

\begin{figure}[htbp]
\centering
\resizebox{0.48\textwidth}{!}{%
\includegraphics{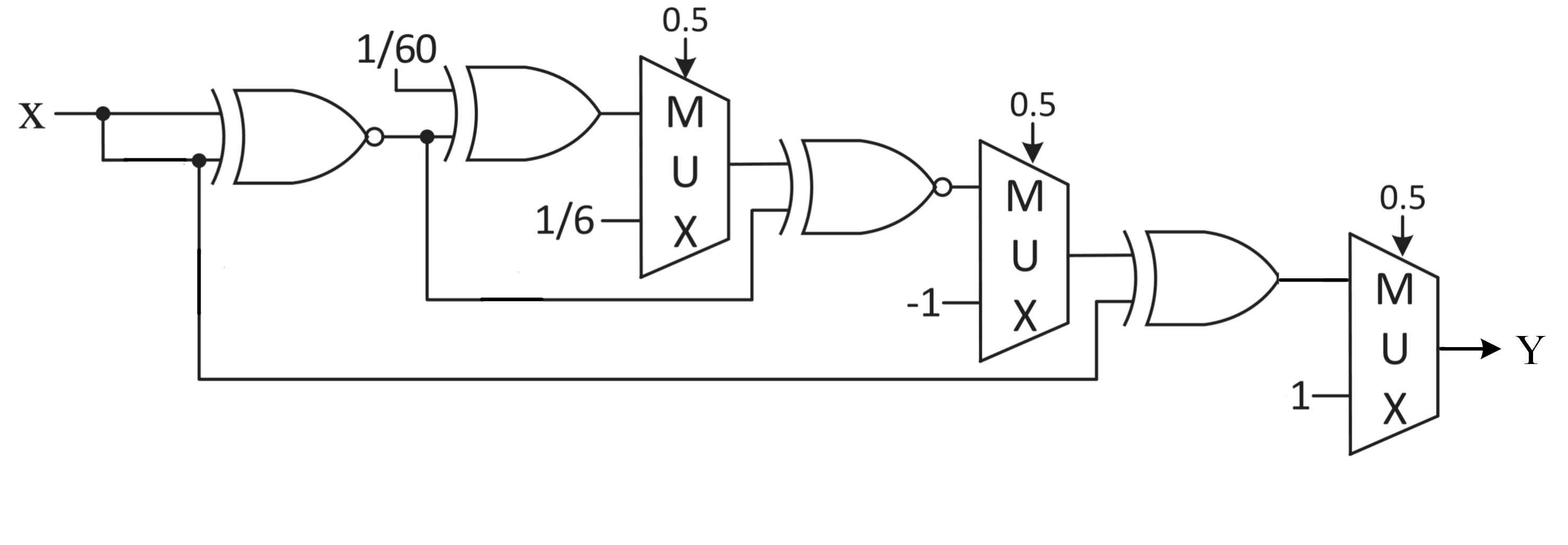}}
\caption{Molecular implementation of $sigmoid(x)$ using $5^{th}$-order Maclaurin expansion and Horner's rule.}
\label{sig}
\end{figure}

Consider the approximation of the sigmoid function given by~\cite{parhi2016computing}:

\begin{eqnarray}
sigmoid(x)&=&\frac{1}{1+e^{-x}} \nonumber\\
&\approx&\frac{1}{2}+\frac{x}{4}+\frac{x^3}{48}+\frac{x^5}{480}\nonumber\\
&=&\frac{1}{2}-\frac{1}{2}x(\frac{1}{2}(-1+x^2\frac{1}{2}(\frac{1}{6}-\frac{x^2}{60})))\label{eqnsig}.
\end{eqnarray}
where $sigmoid(x)$ is approximated by a $5^{th}$-order truncated Maclaurin series and then reformulated by Horner's rule~\cite{parhi1999}. In equation (\ref{eqnsig}), all coefficients, $\frac{1}{2}$, $-\frac{1}{2}$, $-1$, $\frac{1}{6}$ and $-\frac{1}{60}$, can be represented using bipolar format. Fig.~\ref{sig} shows the molecular implementation of $sigmoid(x)$ by cascading \texttt{XOR}, \texttt{XNOR} and \texttt{MUX} gates. The bipolar \texttt{Mult}, \texttt{NMult} and \texttt{MUX} units discussed before compute the same operations as \texttt{XNOR}, \texttt{XOR} and \texttt{MUX} in stochastic implementation, respectively. So equation (\ref{eqnsig}) can be implemented using bipolar \texttt{Mult}, \texttt{NMult} and \texttt{MUX} units shown in Figs.~\ref{base} (c), (d) and (e), respectively.

\begin{figure}[htbp]
\centering
\resizebox{0.40\textwidth}{!}{%
\includegraphics{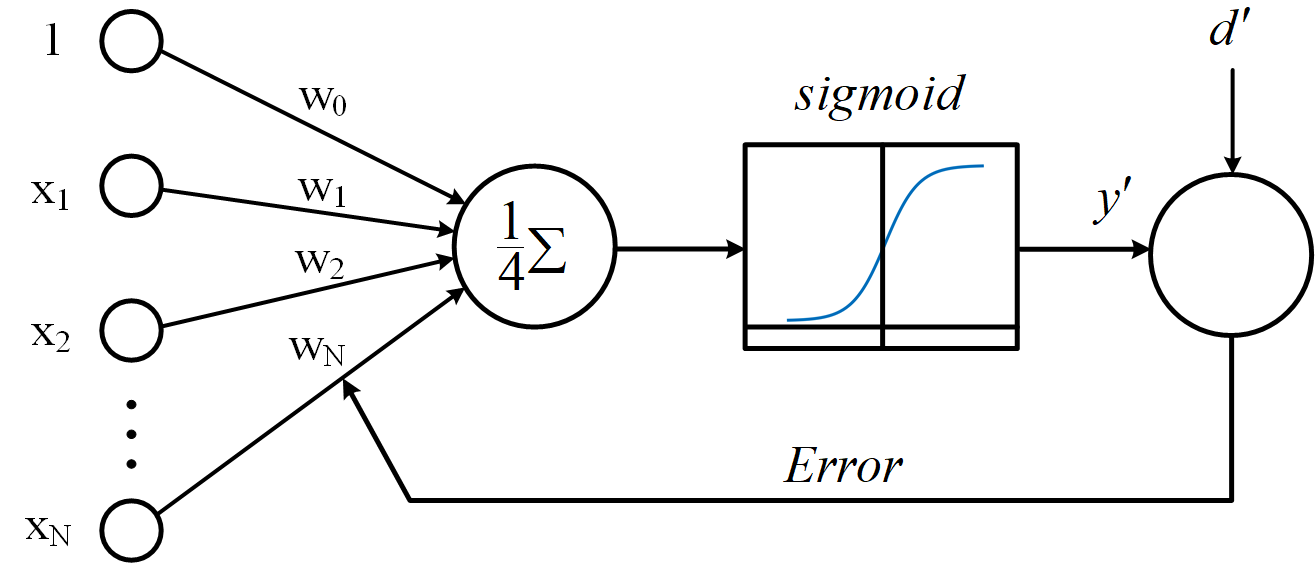}}
\caption{A modified perceptron that can apply sigmoid function to the scaled weighted sum and update weights according to the distance between actual output $y'$ and desired output $d'$.}
\label{modper}
\end{figure}

By cascading the inner product and sigmoid function, the final output of the molecular perceptron is given by $sigmoid(\frac{1}{N}\sum_{i=1}^N w_ix_i)$. Fig.~\ref{modper} shows the modified perceptron where the input to the activation function is the scaled weighted sum, as opposed to the weighted sum. During the training of the molecular perceptron, the error function will be different from the original error function. Here the output of the sigmoid function is different from the desired value. The desired output $d$ of a typical perceptron is modified to $d'$ that can be computed {\em a priori} using:

\begin{eqnarray}
d'&=&sigmoid(\frac{1}{N} logit(d)) \nonumber\\
&=&sigmoid(\frac{1}{N} log(\frac{d}{1-d})) \nonumber \\
&=&\frac{1}{1+(\frac{d}{1-d})^{-\frac{1}{N}}}\label{dp}.
\end{eqnarray}
where $logit$ is the inverse of the $sigmoid$ function and $log$ represents the natural logarithm with base $e$.

\section{Molecular Reactions for Backpropagation in Perceptron}
This section presents the molecular implementation of updating weights in the modified perceptron after forward computation.

Define the loss function $E$ as:

\begin{equation}
\begin{split}
E=\frac{1}{2}(y'-d')^2 \text{ where } y'=sigmoid(\frac{1}{N}\sum_{i=1}^{N}w_ix_i)\nonumber
\end{split}
\end{equation}
represents the actual output of the modified perceptron for a given input vector and $d'$ is the corresponding desired output computed by equation (\ref{dp}). Then the negative of the derivative of the loss in terms of the weights, $-\frac{\partial E}{\partial w_i}$, can be computed as:

\begin{eqnarray}
-\frac{\partial E}{\partial w_i}&=&-\frac{\partial}{\partial w_i} (\frac{1}{2}(y'-d')^2) \nonumber\\
&=&(d'-y')\frac{\partial y'}{\partial w_i} \nonumber\\
&=&(d'-y')\frac{\partial}{\partial w_i} (sigmoid(\frac{1}{N}\sum_{i=1}^{N}w_ix_i)) \nonumber\\
&=&\frac{1}{N}(d'-y')y'(1-y')x_i. \label{delta}
\end{eqnarray}

For molecular implementation, equation (\ref{delta}) can be reformulated as:

\begin{eqnarray}
-\frac{\partial E}{\partial w_i}&=&\frac{4}{N}\frac{1}{2}(d'-y')\frac{1}{2}(y'-y'^2)x_i \nonumber
\end{eqnarray}
where $\frac{4}{N}$ is a constant and can be integrated into the training rate, i.e., by replacing $\alpha$ by  $ \frac{\alpha N}{4}$. Then the update rule for each weight is illustrated as follows:

\begin{eqnarray}
w_{i,new}&=&w_i-\alpha \frac{\partial E}{\partial w_i} \nonumber\\
&=&w_i+\frac{1}{2}(d'-y')\frac{1}{2}(y'-y'^2)x_i \nonumber \\
&=&w_i+\Delta w_i. \label{eqndw}
\end{eqnarray}
where $\Delta w_i=\frac{1}{2}(d'-y')\frac{1}{2}(y'-{y'}^2)x_i$.

\begin{figure}[htbp]
\centering
\resizebox{0.48\textwidth}{!}{%
\includegraphics{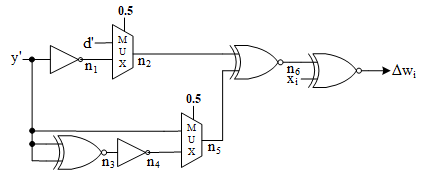}}
\caption{Molecular implementation of $\Delta w_i$.}
\label{dedw}
\end{figure}

The molecular implementation of $\Delta w_i$ in equation (\ref{eqndw}) is shown in Fig.~\ref{dedw}. Note that the input and all coefficients are represented in bipolar format. The internal nodes and final output are described by:

\begin{align}
n_1&=-y',\quad n_2=0.5(d'+n_1),\quad n_3={y'}^2,\quad n_4=-n_3 \nonumber\\
n_5&=0.5(y'+n_4),\quad n_6=n_2n_5,\quad \Delta w_i=n_6x_i \nonumber
\end{align}

Define a modified scaling function as:

\[
y=f(x,M) = 
\begin{dcases*}
   -1,                         & $Mx<-1$\\
   Mx,                         & $-1\leq Mx \leq 1$ \\
   1,                          & $Mx>1$
\end{dcases*}
\]
where input $x$ and output $y$ are in bipolar format and $M\geq 1$. Assume that $[X_1]$ and $[X_0]$ are the concentrations of the two assigned molecular types of the input $x$ of the modified scaling function in bipolar format. Also assume that $[Y_1^+]$ and $[Y_0^+]$ are the concentrations of the two assigned molecular types of the output $y$ of the modified scaling function. The relationship between the two sets of concentrations can be written as:

\begin{align}
y = \frac{[Y_1^+]-[Y_0^+]}{[Y_1^+]+[Y_0^+]}=M\frac{[X_1]-[X_0]}{[X_1]+[X_0]}=Mx. \label{convert}
\end{align}

The modified scaling function described by equation (\ref{convert}) can be realized by the following molecular reactions:

\begin{align}
&X_0\longrightarrow (M+1)Y_0^+ +(M-1)Y_1^-\nonumber\\
&X_1\longrightarrow (M+1)Y_1^+ +(M-1)Y_0^-\nonumber\\
&Y_0^+ + Y_0^- \xrightarrow{fast} \text{\O} \nonumber\\
&Y_1^+ + Y_1^- \xrightarrow{fast} \text{\O}
\label{con}
\end{align}

\begin{figure}[htbp]
\centering
\resizebox{0.30\textwidth}{!}{%
\includegraphics{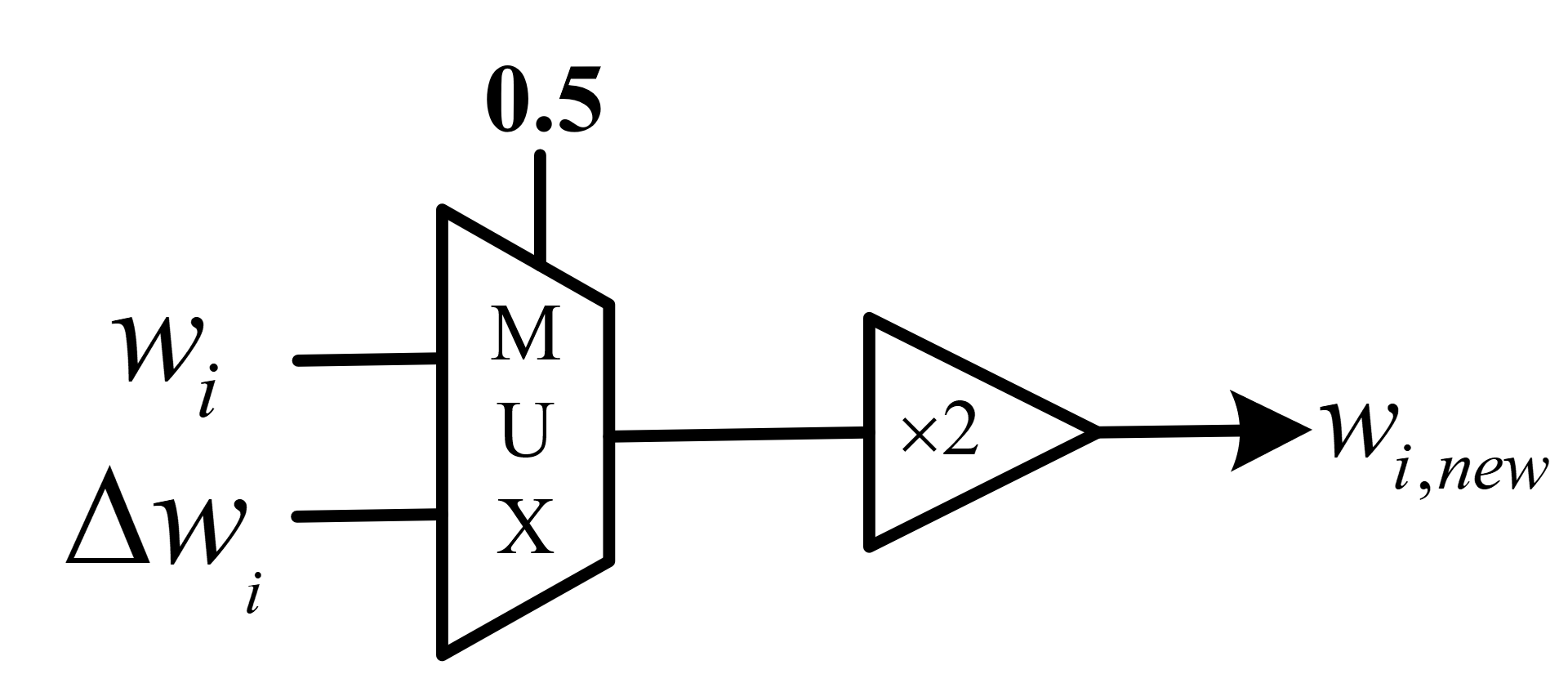}}
\caption{Molecular implementation of $w_{i,new}$ with two inputs $w_i$ and $\Delta w_i$.}
\label{update}
\end{figure}

After the eight molecular reactions shown in equation (\ref{con}) are completed, the molecules $Y_1^+$ and $Y_0^+$ are treated as $1$-molecule and $0$-molecule of $y$, respectively. Then, $y = \frac{[Y_1^+]-[Y_0^+]}{[Y_1^+]+[Y_0^+]}$ is the output of the modified scaleing function. Therefore, to compute $w_{i,new}$ as required in equation (\ref{eqndw}), we can reformulate this equation into equation (\ref{eqndwr}) and implement it by cascading a \texttt{MUX} unit with two inputs ($w_i$ and $\Delta w_i$) and a modified scaling function with scale factor $M=2$ as shown in Fig.~\ref{update}.

\begin{eqnarray}
w_{i,new}=2\times \frac{1}{2}(w_i+\Delta w_i).\label{eqndwr}
\end{eqnarray}

\section{Evaluation of Perceptron using DNA}
Abstract chemical reaction networks (CRNs) described by molecular reactions can be mapped to DNA strand displacement (DSD) reactions as shown in \cite{soloveichik2010dna}. A framework that can implement arbitrary molecular reactions with no more than two reactants by linear, double-stranded DNA complexes was proposed in~\cite{soloveichik2010dna}. Notice that our computational units are all built based on molecular reactions with at most two reactants. We simulate the perceptron that can learn by using the software package provided by Winfree's team at Caltech \cite{soloveichik2010dna}. More details of mapping bimolecular reactions to DSD are also described in \cite{soloveichik2010dna}.

\begin{figure}[htbp]
\centering
\resizebox{0.48\textwidth}{!}{%
\includegraphics{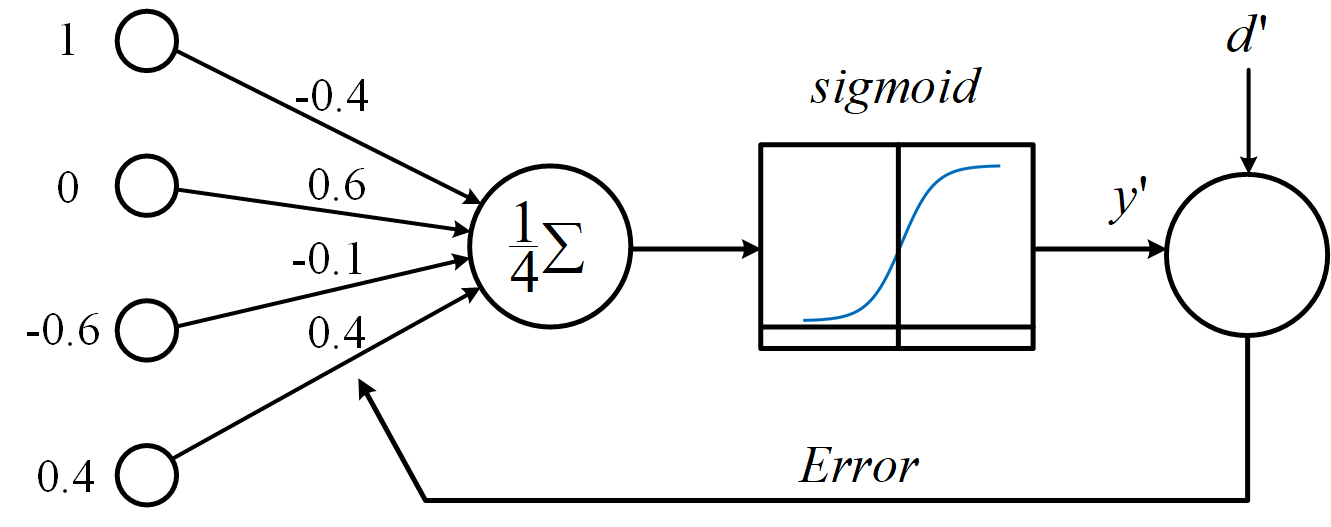}}
\caption{Evaluated modified perceptron with $3$ inputs and $1$ bias.}
\label{perceptron}
\end{figure}

A simple modified perceptron with $3$ inputs and $1$ bias is evaluated as shown in Fig.~\ref{perceptron}. Two sets of input-output pairs are selected to demonstrate the functionality of the modified perceptron that can learn. The input vector, initial weight vector and bias of both sets are $\begin{bmatrix}0\\-0.6\\0.4\end{bmatrix}$, $\begin{bmatrix}0.6\\-0.1\\0.4\end{bmatrix}$ and $-0.4$, respectively. The only difference between these two data sets is the desired output value. Given $d=0.835$ and $0.309$ for these two data sets, we can get $d'=0.6$ and $0.45$ for these two data sets by using equation (\ref{dp}). In a real application, one input-output pair from the whole data set should be fed into the perceptron during each epoch. To show the convergence of the molecular perceptron, we input the same input-output pair to the modified perceptron each epoch. The simple modified perceptron with forward computation and backpropagation is mapped to DNA using DSD.

\begin{figure}[htbp]
\centering
\resizebox{0.4\textwidth}{!}{%
\includegraphics{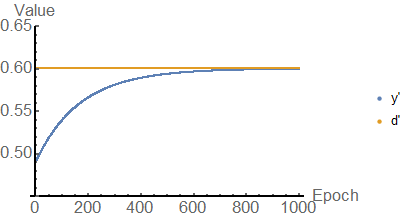}}
\caption{Forward computation result of the modified perceptron during training with desired output $d'=0.6$.}
\label{06}
\end{figure}

\begin{figure}[htbp]
\centering
\resizebox{0.4\textwidth}{!}{%
\includegraphics{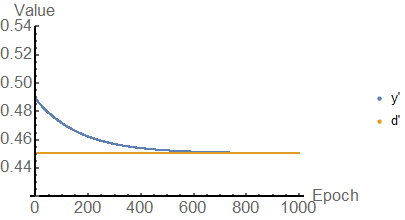}}
\caption{Forward computation result of the modified perceptron during training with desired output $d'=0.45$.}
\label{045}
\end{figure}

\begin{figure}[htbp]
\centering
\resizebox{0.4\textwidth}{!}{%
\includegraphics{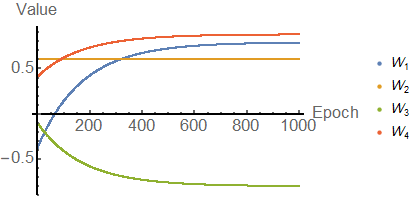}}
\caption{Weights and bias of the modified perceptron during training with desired output $d'=0.6$.}
\label{06w}
\end{figure}

\begin{figure}[htbp]
\centering
\resizebox{0.4\textwidth}{!}{%
\includegraphics{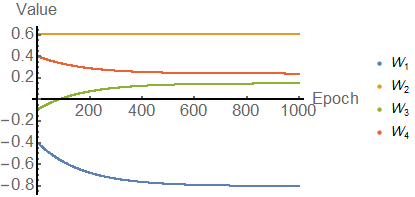}}
\caption{Weights and bias of the modified perceptron during training with desired output $d'=0.45$.}
\label{045w}
\end{figure}

Figs.~\ref{06} and \ref{045} show the forward computation results of the modified perceptron during training with desired outputs $0.6$ and $0.45$, respectively. The yellow lines illustrate the desired outputs, and the blue lines illustrate the convergence of the output. Figs.~\ref{06w} and \ref{045w} illustrate the convergence of weights and biases during training. The blue, yellow, green and red lines represent the values of $w_1$, $w_2$, $w_3$ and $w_4$, respectively. The horizontal axis in these four figures represents the epoch index.

\section{Conclusion}
This paper has presented the molecular implementations of backpropagation in a perceptron using DNA. A molecular perceptron with a rectified linear unit (ReLU) can be synthesized in a similar manner using the molecular ReLU function described in \cite{liu2019molecular}. The delay element for weight update has not been integrated into the molecular perceptron. However, the molecular delay element can be realized using either an asynchronous RGB clock or a synchronous clock \cite{jiang2012digital,jiang2013discrete,jiang2011synchronous}.

\ifCLASSOPTIONcaptionsoff
  \newpage
\fi

\bibliographystyle{ieeetr} 
\bibliography{main}

\end{document}